\documentclass[conference]{IEEEtran}
\IEEEoverridecommandlockouts
\usepackage{cite}
\usepackage{amsmath,amssymb,amsfonts}
\usepackage{algorithmic}
\usepackage[ruled,linesnumbered]{algorithm2e}
\usepackage{array}
\usepackage{multirow}
\usepackage[caption=false,font=normalsize,labelfont=sf,textfont=sf]{subfig}
\usepackage{textcomp}
\usepackage{stfloats}
\usepackage{url}
\usepackage{verbatim}
\usepackage{graphicx}
\usepackage{cite}
\usepackage{makecell}
\usepackage{textcomp}
\usepackage{xcolor}
\usepackage{balance}
\usepackage{float}
\usepackage[switch]{lineno}

\usepackage{graphicx}
\usepackage{textcomp}
\usepackage{xcolor}
\def\BibTeX{{\rm B\kern-.05em{\sc i\kern-.025em b}\kern-.08em
    T\kern-.1667em\lower.7ex\hbox{E}\kern-.125emX}}
\begin{document}
\title{Large Scale Finite-Temperature Real-time Time Dependent Density Functional Theory Calculation with Hybrid Functional on ARM and GPU Systems}

\author{
\IEEEauthorblockN{Rongrong Liu}
\IEEEauthorblockA{
\textit{State Key Lab of Processors,}\\ 
\textit{Institute of Computing Technology,}\\ 
\textit{Chinese Academy of Sciences,}\\
\textit{University of Chinese Academy}\\
\textit{of Sciences,}
Beijing, China \\
liurongrong21s@ict.ac.cn}\\
\IEEEauthorblockN{Tong Zhao}
\IEEEauthorblockA{
\textit{State Key Lab of Processors,}\\ 
\textit{Institute of Computing Technology,}\\ 
\textit{Chinese Academy of Sciences,}\\
\textit{University of Chinese Academy}\\
\textit{of Sciences,}
Beijing, China \\
zhaotong@ict.ac.cn}\\
\IEEEauthorblockN{Lijun Liu}
\IEEEauthorblockA{
\textit{Department of Mechanical Engineering,} \\
\textit{Graduate School of Engineering,} \\
\textit{Osaka University,}\\
Osaka, Japan\\
liu@mech.eng.osaka-u.ac.jp}
\and
\IEEEauthorblockN{Zhuoqiang Guo}
\IEEEauthorblockA{
\textit{State Key Lab of Processors,}\\ 
\textit{Institute of Computing Technology,}\\ 
\textit{Chinese Academy of Sciences,}\\
\textit{University of Chinese Academy}\\
\textit{of Sciences,}
Beijing, China \\
guozhuoqiang20z@ict.ac.cn}\\
\IEEEauthorblockN{Haibo Li}
\IEEEauthorblockA{
\textit{School of Mathematics and }\\
\textit{Statistics, The University of} \\
\textit{Melbourne,}
 Melbourne, Australia \\
haibo.li@unimelb.edu.au}\\\\\\
\IEEEauthorblockN{Guangming Tan}
\IEEEauthorblockA{
\textit{State Key Lab of Processors,}\\ 
\textit{Institute of Computing Technology,}\\ 
\textit{Chinese Academy of Sciences,}\\
\textit{University of Chinese Academy}\\
\textit{of Sciences,}
Beijing, China \\
tgm@ict.ac.cn}
\and
\IEEEauthorblockN{Qiuchen Sha}
\IEEEauthorblockA{
\textit{State Key Lab of Processors,}\\ 
\textit{Institute of Computing Technology,}\\ 
\textit{Chinese Academy of Sciences,}\\
\textit{University of Chinese Academy}\\
\textit{of Sciences,}
Beijing, China \\
shaqiuchen22s@ict.ac.cn}\\
\IEEEauthorblockN{WeiHu}
\IEEEauthorblockA{
\textit{School of Computer Science}\\
\textit{and Technology, University of}\\
\textit{Science and Technology of }\\
\textit{China, }
Hefei, China \\
whuustc@ustc.edu.cn}\\\\
\IEEEauthorblockN{Weile Jia}
\IEEEauthorblockA{
\textit{State Key Lab of Processors,}\\ 
\textit{Institute of Computing Technology,}\\ 
\textit{Chinese Academy of Sciences,}\\
\textit{University of Chinese Academy}\\
\textit{of Sciences,}
Beijing, China \\
jiaweile@ict.ac.cn}
}


\maketitle

\begin{abstract}
Ultra-fast electronic phenomena originating from finite temperature, such as nonlinear optical excitation, can be simulated with high fidelity via real-time time dependent density functional theory (rt-TDDFT) calculations  with hybrid functional. However, previous rt-TDDFT simulations of real materials using the optimal gauge--known as the parallel transport gauge--have been limited to low-temperature systems with band gaps. In this paper, we introduce the parallel transport-implicit midpoint (PT-IM) method, which  significantly accelerates finite-temperature rt-TDDFT calculations of real materials with hybrid function. We first implement PT-IM with hybrid functional in our plane wave code PWDFT, and optimized it on both GPU and ARM platforms to build a solid baseline code. Next, we propose a diagonalization method to reduce computation and communication complexity, and then, we employ adaptively compressed exchange (ACE) method to reduce the frequency of the most expensive Fock exchange operator. Finally, we adopt the ring\_based method and the shared memory mechanism to overlap computation and communication and alleviate memory consumption respectively. Numerical results show that our optimized code can reach 3072 atoms for rt-TDDFT simulation with hybrid functional at finite temperature on 192 computing nodes, the time-to-solution for one time step is 429.3s, which is 41.4 times faster compared to the baseline.
\end{abstract}


\section{Introduction}
Real-time time-dependent density functional theory (rt-TDDFT)~\cite{andrade2012time,onida2002electronic,runge1984density,ullrich2011time,wu2009order} is a widely used approach in electronic excitation calculations, gaining research attention with the growing experimental focus on ultrafast electronic phenomena in materials science. It can be used in a spectrum of applications, including ion collisions~\cite{wang2015efficient}, the light absorption spectrum~\cite{fischer2015excited}, laser-induced demagnetization and phase transitions~\cite{liu2020microscopic}, charge transfer, dynamics of excited carriers, and chemical reactions \cite{casida2012progress}. Recent studies ~\cite{cushing2017plasmonic,lucchini2016attosecond,moulet2017soft,schlaepfer2018attosecond,schultze2014attosecond,tan2017plasmonic,zurch2017direct} have illuminated the capacity of laser excitation to initiate structural phase transitions and charge density wave excitations, along with revealing that many interactions in catalysis, previously assumed to be adiabatic, actually proceed via non-adiabatic mechanisms, necessitating electron excitation simulations for accurate analysis. These developments mark a significant shift in materials science simulations, pushing the boundaries beyond conventional ground state DFT calculation.

However, rt-TDDFT simulations of real materials still face two challenges: first, rt-TDDFT calculations are constrained by the precision and stability requisites of ordinary differential equation (ODE) integrators, restricting the time step to the sub-attosecond domain to ensure the accuracy of dynamics. Consequently, explicit time integrators, notably the fourth-order Runge-Kutta (RK4) method, are frequently favored over implicit methods like the Crank-Nicolson (CN) scheme, owing to their operational efficiency and ease of implementation. The parallel transport gauge formalism can find the slowest oscillating orbitals, allowing for the effective utilization of implicit integrators with significantly larger step sizes. The parallel transport Crank-Nicolson (PT-CN) scheme, in particular, has been shown to extend the feasible time step to approximately 50 attoseconds while maintaining accuracy comparable to the RK4 method~\cite{jia2018fast}. 
However, the current PT-CN scheme is only applicable for systems with band gaps, which means PT-CN cannot be applied to metallic or finite-temperature systems where electrons are fractionally occupied. Secondly, semi-local exchange-correlation functionals, like the Local Density Approximation (LDA)~\cite{ceperley1980ground,perdew1981self} and Generalized Gradient Approximation (GGA)~\cite{perdew1996generalized}, fall short in accurately describing excited states and band gaps, which can lead to incorrect behavior such as the emergence of nonphysical exciton in rt-TDDFT calculations. 
Hybrid  functionals~\cite{raghavachari2000perspective,perdew1996rationale} within Density Functional Theory (DFT) offer a solution by mixing a portion of the Fock exchange integral with semi-local functionals, improving accuracy for electronic structures. Particularly, range-separated hybrid functionals~\cite{heyd2003hybrid,heyb2006erratum} have shown to match optical absorption spectra accurately, comparable to results from more demanding methods like the Bethe-Salpeter equation based on GW calculations~\cite{del2020accelerating}. Thus, combining rt-TDDFT with hybrid functionals represents a powerful approach for accurately modeling exciton excitation and charge transfer, marking a notable advance in materials science simulations. 

Unfortunately, both finite-temperature rt-TDDFT and hybrid functional require significantly higher computational costs compared to traditional ground state DFT calculations with semi-local functionals. Specifically, rt-TDDFT can be orders of magnitude slower than conventional ground state molecular dynamics simulations, attributed to its smaller time steps. Similarly, hybrid functionals are substantially slower than semi-local exchange-correlation functionals due to the evalution of the Fock exchange term. Consequently, literature on plane wave-based real material rt-TDDFT simulation with hybrid functional, even for modest systems comprising a few atoms, is scarce, let alone for larger systems containing thousands of atoms~\cite{jia2019parallel,alducin2017non}. Nonetheless, for a multitude of applications, such as excited state charge transfer and laser-induced structural phase transitions in nanostructures like quantum dots and wires, simulating large systems is indispensable. Moreover, complete basis sets such as plane waves are more favorable in describing excited states in rt-TDDFT calculations. All these considerations pose challenges to rt-TDDFT calculations of real materials at finite temperatures. 

The latest developments in high-performance computing and the parallel transport gauge formalism have offers opportunities for overcoming these challenges, both from algorithmic and hardware perspectives. Recently, An et al. proposed a parallel transport formalism for rt-TDDFT at finite temperatures~\cite{an2022parallel}, demonstrating its potential of extending the time step length significantly beyond the limitations of the Runge-Kutta 4th order (RK4) method in a one-dimension problem. On the hardware side, with the improvement of architecture and manufacturing processes, the computing performance has been enhanced at a faster pace compared to that of memory bandwidth and network bandwidth, resulting in an increasingly wider gap between the former and the latter two.
This raises a highly intriguing question: can hybrid functional rt-TDDFT calculations be accelerated on many-core systems, such as ARM or GPU platforms?

In this paper, we present our highly scalable and efficient implementation of finite-temperature rt-TDDFT calculation with hybrid functionals, and optimized for both ARM and GPU platforms using the planewave code PWDFT. Our major contributions are as follows:
\begin{itemize}
    \item We implement finite-temperature rt-TDDFT algorithm PT-IM with hybrid functional for 3D real materials and optimized it on ARM and GPU platforms using OpenMP and CUDA, respectively.
    \item We further proposed a matrix diagonalization method to reduce the computational complexity from $O(N^4)$ to $O(N^3)$ and significantly decreased the frequency of the most expensive hybrid functional calculations using the  Adaptively Compressed Exchange (ACE) method.
    \item Additionally, we proposed an Asynchronous ring-based method and utilized a shared memory mechanism to optimize the network communication and reduce memory consumption.
    \item Testing results of a 384-atom silicon system show that compared to the baseline, our optimized code achieves a speedup of 55.15 and 41.44 times on ARM and GPU platforms, respectively. Our optimized code can also scale up to 960 nodes on Fugaku (46080 ARM cores) to simulate a silicon system of 1536 atoms and can reach 3072 atoms (12288 electrons) on 768 A100 GPUs. 
\end{itemize}
This paper is organized as follows: we review the rt-TDDFT algorithm PT-IM in Sec.~\ref{sec:background}. Then a baseline implementation of PT-IM with hybrid functional is shown in Sec.~\ref{sec:implementation}. We further optimize the PT-IM in Sec.~\ref{sec:further_optimization}. Machine configuration and physical systems are listed in Sec.~\ref{sec:machine} and Sec.~\ref{sec:physical_system}. The physical and performance results are shown in Sec.~\ref{sec:physical_results} and Sec.~\ref{sec:performance_results}, respectively. The conclusion is drawn in Sec.~\ref{sec:conclusion}.

\section{Background}\label{sec:background}
\subsection{The parallel transport-implicit midpoint (PT-IM) method}\label{sec:ptim}
Real-time time-dependent density functional theory solves the following time-dependent equation:
\begin{equation}
i\partial_{t}\Psi(t)=H(t,P(t))\Psi(t). \label{eqn:schrodinger}
\end{equation}
Here $\Psi(t)=[\psi_{1}(t),...,\psi_{N}(t)]$ is the collection of electron wavefunctions (also called electron orbitals), and $N$ is the number of total electron states (spin degeneracy omitted). $P(t)$ is the density matrix, which defined as $P(t) =\Psi(t)\sigma(t)\Psi^{\ast}(t)$\cite{an2022parallel}. $\Psi^{\ast}$ is the Hermitian conjugate of $\Psi$ and $\sigma(t)$ is the occupation number matrix.

In pure states (low temperature), $\sigma(t) = \sigma(0) = I_N$. So $P(r,r') = \sum\limits_{i}^{N}\psi_{i}(r)\sigma_{i}\psi^{\ast}_{i}(r')$, and $\sigma_{i}$ is either one (occupied) or zero (unoccupied). 
In real material rt-TDDFT simulation, the initial state $\sigma(0)$ is required to be a mixed state. For instance, for metallic systems or semiconductors at finite temperatures, the wavefunctions can be fractionally occupied by the Fermi-Dirac distribution. In such mixed states, 
\begin{equation}
P(r,r') = \sum\limits_{i,j=1}^{N}\psi_{i}(r)\sigma_{ij}\psi^{\ast}_{j}(r').\label{eqn:mix_P(r,r')}
\end{equation}

The corresponding rt-TDDFT equation~\ref{eqn:schrodinger} can be equivalently reformulated using a series of unitarily transformed orbitals. Physical observables, including the density matrix, remain unchanged under such unitary transformations, a property known as gauge invariance. This invariance enables the pursuit of an optimal gauge. Recent advancements have pinpointed such an optimal gauge \cite{an2022parallel}, implicitly defined by the subsequent equation:
\begin{equation}
\begin{split}
i\partial_{t} \Phi(t) &=  (I - \Tilde{P}(t))H(t,P(t))\Phi(t),\\
i\partial_{t} \sigma(t) &= [(\Phi^{\ast}(t)H(t,P(t))\Phi(t), \sigma(t)],\label{eqn:PTIM}
\end{split}
\end{equation}
where $\Phi(t)$ oscillates much slower by choosing the optimal gauge ($\Phi(t)=\Psi(t)U(t)$).
Coupled with the implicit midpoint (IM) rule (also known as the Gauss-Legendre method of order 2), the shorthand notations are introduced:
\begin{equation}
\Phi_{n+\frac{1}{2}} =\frac{\Phi_{n+1} + \Phi_{n}}{2}, \sigma_{n+\frac{1}{2}} =\frac{\sigma_{n+1} + \sigma_{n}}{2},\label{eqn:n+1/2}
\end{equation}
and accordingly,
\begin{equation}
\begin{split}
\Tilde{P}_{n+\frac{1}{2}} &=\Phi_{n+\frac{1}{2}}(\Phi^{\ast}_{n+\frac{1}{2}}\Phi_{n+\frac{1}{2}})^{-1}\Phi^{\ast}_{n+\frac{1}{2}},\\
{P}_{n+\frac{1}{2}} &=\Phi_{n+\frac{1}{2}}\sigma_{n+\frac{1}{2}}\Phi^{\ast}_{n+\frac{1}{2}},\\
H_{n+\frac{1}{2}} &= H(t_{n+\frac{1}{2}},P_{n+\frac{1}{2}}),\label{eqn:p_tilde}
\end{split}
\end{equation}
the parallel transport-implicit midpoint scheme (PT-IM) at each time step reads:
\begin{equation}
\begin{split}
\Phi_{n+1} &=  \Phi_{n} - i\Delta_{t}(I - \Tilde{P}_{n+\frac{1}{2}})H_{n+\frac{1}{2}}\Phi_{n+\frac{1}{2}}),\\
\sigma_{n+1} &=  \sigma_{n} - i\Delta_{t}[(\Phi^{\ast}_{n+\frac{1}{2}}H_{n+\frac{1}{2}}\Phi_{n+\frac{1}{2}}), \sigma_{n+\frac{1}{2}}].\label{eqn:lisan}
\end{split}
\end{equation}
If $\{\Phi_{n+1};\sigma_{n+1}\}$ is chosen to be the unknowns, then equation \ref{eqn:PTIM} can be viewed as a fixed point equation in the abstract form
\begin{equation}
x = T(x).\label{eqn:x=tx}
\end{equation}

\subsection{Fock exchange operator}\label{sec:hybrid_functional}
The Hamiltonian has the following operators when hybrid functionals are used:
\begin{equation}
H[P] = -\frac{1}{2}\Delta + V_{ext}(t) + V_{Hxc}[P(t)] + \alpha V_{x}[P(t)]. \label{1}
\end{equation}

Here $V_{ext}(t)$ is the time-dependent external potential and $V_{Hxc}$ consists of the Hartree potential and the local part of the exchange-correlation potential. Without the term $V_{x}$,  the functional is considered semilocal. This paper focuses on hybrid functional rt-TDDFT calculations, where $V_{x}$, called the Fock exchange operator, is an integral operator with kernel $V_{x}[P](r,r') = -P(r,r')K(r,r')$. In this context, $K(r,r')$ denotes the kernel for the (possibly screened) electron interaction, and $\alpha$ represents a mixing fraction (usually $\alpha = 0.25$).
 
In hybrid functional calculations, in pure states, each set of multiplications $V_{x}[P]\Phi$ requires the following operations:
\begin{equation}
\begin{split}
(V_{x}[P]\phi_{j})(r) = -\sum\limits_{i=1}^{N}\phi_{i}(r)\sigma_{i}\int K(r,r')\phi^{\ast}_{i}(r')\phi_{j}(r')dr'. \label{eqn:pure_vx}
\end{split}
\end{equation}
In mixed states at finite temperatures, with $P(r,r')$ in the form(\ref{eqn:mix_P(r,r')}), each set of multiplications $V_{x}[P]\Phi$ takes the following form:
\begin{equation}
(V_{x}[P])\phi_{j}(r)= -\sum\limits_{i,k=1}^{N}\sigma_{ik}\phi_{i}(r)\int K(r,r')\phi^{\ast}_{k}(r')\phi_{j}(r')dr'.\label{eqn:mixed_vx}
\end{equation}
In planewave basis, for $V_x$ applied to a single orbital, in pure states, it can be calculated via solving $N^2$ Poisson-type equations. In mixed states, however, it amounts to solving $N^3$ Poisson equations. And then we need to do this for all $N$ orbitals to obtain $V_x\Phi$. If we denote the number of discrete lattice points in real space by ${N_ {g}}$, in pure states, the total cost of $V_x\Phi$ is $O(N_{g}logN_{g}N^2) \sim O(N^3)$ and in mixed states the cost is $O(N_{g}logN_{g}N^{3}) \sim O(N^4)$. Notably, in pure states, the time taken by the Fock exchange operator already accounts for over 95\% of the total time, meaning that, for the same quantum system, the solution time of hybrid functional DFT is more than 20 times that of semi-local functional DFT \cite{jia2019parallel}. In mixed states, the computational complexity of the Fock exchange operator increases by an order. This implies that performing rt-TDDFT with hybrid functionals on large systems at finite temperatures is prohibitively expensive.

\section{Baseline implementation of PT-IM with hybrid functional on GPU and ARM platforms}~\label{sec:implementation}
Since there is no prior plane-wave implementation of the PT-IM method, the primary task of this paper is to develop an efficient baseline version of PT-IM. This section details our approach to implementing an efficient PT-IM method on GPU and ARM platforms using MPI combined with OpenMP/CUDA in the PWDFT package~\cite{jia2019parallel}.
Note that OpenMP and GPU acceleration have been adopted for ground state electronic structure calculations in several software packages, including ABINIT~\cite{gonze2016recent}, 
PWmat~\cite{jia2013analysis,jia2013fast}, Quantum ESPRESSO~\cite{romero2018performance}, VASP~\cite{hacene2012accelerating}, BigDFT ~\cite{ratcliff2018affordable}, NWChem~\cite{valiev2010nwchem}. To achieve a better acceleration, our optimization efforts extended beyond the computationally expensive Fock exchange operator to include acceleration of additional components, such as residual calculations and wavefunction mixing.
\begin{algorithm}[h]
  \SetAlgoLined
  \KwIn{$\Phi_{n}$ and $\sigma_{n}$} 
  \KwOut{$\Phi_{n+1}$ and $\sigma_{n+1}$ }
Suppose $\{\Phi_{n+1},\sigma_{n+1} \} = T(\{\Phi_{n},\sigma_{n}\})$ \;
Calculate $\rho^{in}_{n+1}$ from $\Phi_{n+1}$ and $\sigma_{n+1}$\;
  \For{$k = 1,2...$}{
Calculate $\Phi_{n+\frac{1}{2}}$ and $\sigma_{n+\frac{1}{2}}$ refer to \eqref{eqn:n+1/2}\;
Calculate $\rho_{n+\frac{1}{2}}$ from $\Phi_{n+\frac{1}{2}}$ and  $\sigma_{n+\frac{1}{2}}$\;
Update $H_{n+\frac{1}{2}}$\;
Update $\{\Phi_{n+1},\sigma_{n+1}\}$ refer to \eqref{eqn:lisan}\;
Update $\Phi_{n+1}$ and $\sigma_{n+1}$ by Anderson mixing\;
Evaluate the residual $R_f$ of \eqref{eqn:lisan}\;
Calculate $\rho^{out}_{n+1}$ from $\Phi_{n+1}$ and $\sigma_{n+1}$\;
Jump out of the loop when the density change is sufficiently small\;
}
Orthogonalize $\Phi_{n+1}$ and conjugate symmetrize $\sigma_{n+1}$\;
\caption{One time propagation step for PT-IM method with hybrid functional}
\label{alg:ptim}
\end{algorithm}

Alg.\ref{alg:ptim} outlines a single time propagation step of the PT-IM method. 
First, the initial values of $\Phi_{n+1}$ and $\sigma_{n+1}$ are evaluated to obtain the intermediate wavefunctions $\Phi_{n+\frac{1}{2}}$ and occupation number matrix $\sigma_{n+\frac{1}{2}}$. Next, we calculate the physical quantities at these intermediate moments: $\rho_{n+\frac{1}{2}}$ and $H_{n+\frac{1}{2}}$, which are ultimately used to update the new $\{\Phi_{n+1},\sigma_{n+1}\}$ refer to \eqref{eqn:lisan}, involving the calculation of the Fock exchange operator. Anderson mixing~\cite{anderson1965iterative} of the wavefunctions and charge density are employed to accelerate the convergence of the fixed-point problem.
When the residual of  $\rho$ is sufficiently small, the SCF iteration can be terminated. 
In practice, we find that the SCF convergence can also be controlled by the convergence of the charge density. In Alg.\ref{alg:ptim}, the most time-consuming part is the Fock exchange operator. Therefore, we will focus on optimizing it in the following. Other important computation modules include electron density, residual, and Anderson mixing.
\subsection{Data distribution}\label{sec:paralellization_scheme}
\begin{figure}[ht]
\centering
\scalebox{0.8}{
  \includegraphics[width=0.40\textwidth]{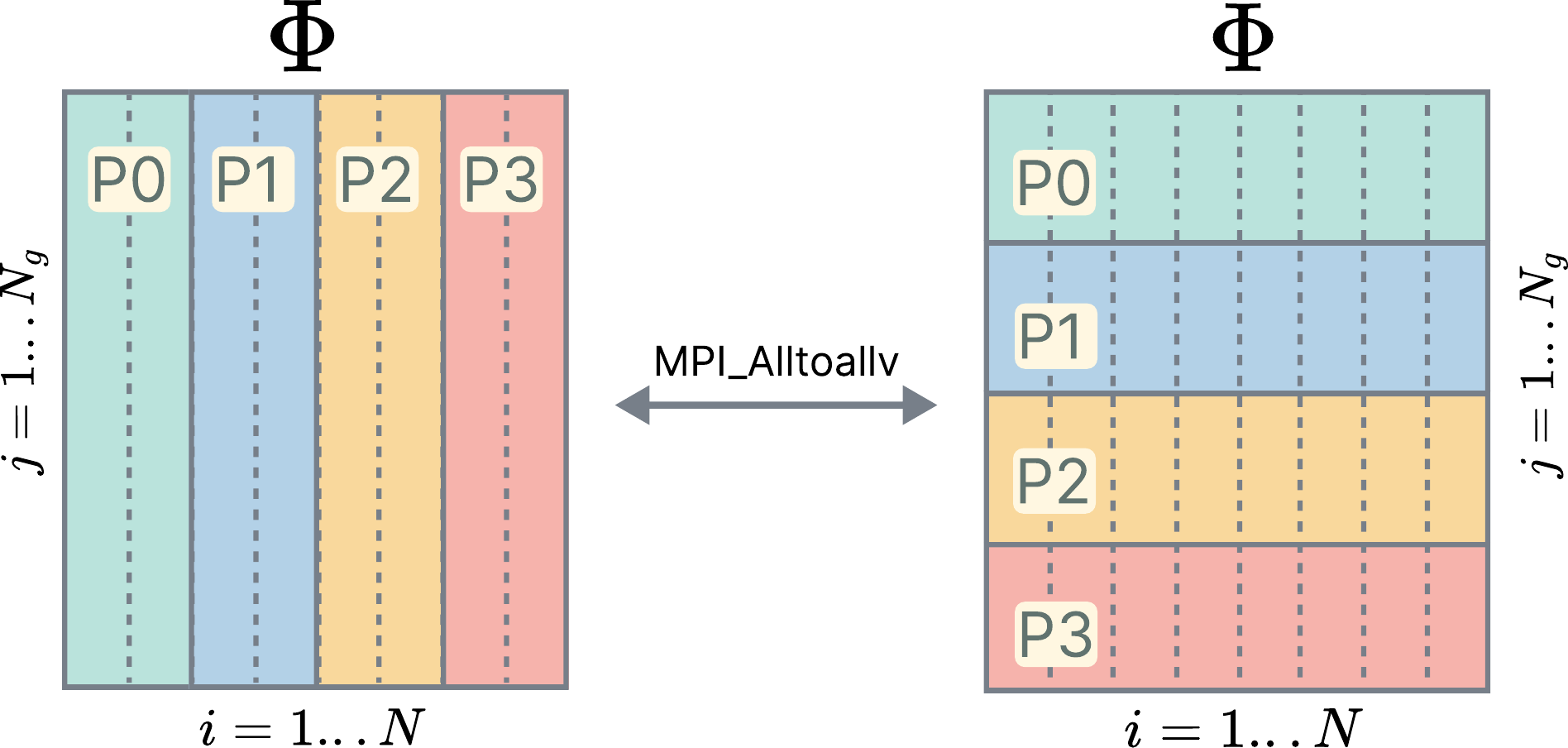}}
  \caption{The parallel distribution of wavefunction $\Phi$ (left: band-index parallelization; right: grid-point parallelization). Note that MPI\_Alltoallv is required to transpose between the two parallelization schemes.}
  \label{fig:parallel}
\end{figure}
Fig.~\ref{fig:parallel} shows the two primary parallelization schemes used in PWDFT. First, the wavefunction $\Phi$ can be distributed over the columns (band-index parallelization as shown in Fig.~\ref{fig:parallel} left). This is particularly efficient for the calculation of $H\Phi$ and hybrid functional since different MPI tasks can perform fast Fourier transformations (FFT) independently. The second parallelization scheme is to distribute the wavefunction $\Phi$ over the rows(grid-point parallelization as shown in Fig.~\ref{fig:parallel} right, G is the grid in Fourier space). grid-point parallelization is efficient for the calculation of the overlap matrix $S=\Phi^{\ast}$H$\Phi$ over matrix-matrix multiplication. Note that MPI\_Alltoall is required to transpose between these two parallelization schemes. And since we focus on large systems with more than a few hundred atoms, only one $\Gamma$ point is needed. Therefore, K-point parallelization is omitted in this paper. 
\subsection{Evaluation of the Fock exchange operator}\label{sec:vx_calculation}
\begin{algorithm}[h]
  \SetAlgoLined
  \KwIn{$\Phi$ and $\sigma$} 
  \KwOut{ $V_x\Phi$ }
Let $V_x\Phi$ be distributed by band-index parallelization and initialized to zero and
$\phi_{temp}$ is a temp variable;\
\For{$k = 1,N$}{
\If{the current process holds $\phi_k$}{
 Broadcast $\phi_k$ to all processes\;
}
\For{$i = 1,N$}{
\If {the current process holds $\phi_i$ }
{Broadcast $\phi_i$ to all processes\;}
\For{$j = 1,N$}{
\If {the current process holds $\phi_j$ }
{
$\phi_{temp} = \phi^{\ast}_{k} \odot \phi_{j}$\;
$\phi_{temp}$ = inplace forward FFT($\phi_{temp}$)\;
$\phi_{temp} = K(r,r') \phi_{temp}$\;
$\phi_{temp}$ = inplace inverse FFT($\phi_{temp}$)\;
$V_{x}\phi_{j} = V_{x}\phi_{j} + \sigma_{ik} \phi_{temp} \odot \phi_{i} $\;
}}}}
\caption{The Fock exchange operator calculation in mixed states}
\label{alg:vx_phi}
\end{algorithm}
In PT-IM implementation, the evaluation of the Fock exchange operator is the most time-consuming part and is repeatedly performed within the matrix-vector multiplication $H\Phi$. Alg.~\ref{alg:vx_phi} details the evaluation of the Fock exchange operator ($V_{x}[P]\Phi$) in mixed-state rt-TDDFT calculation. As discussed earlier, wavefunction $\Phi$ is distributed in band-index parallelization to efficiently perform FFTs. Each wavefunction $\phi_k$ has to be broadcast to all MPI tasks via MPI\_Bcast. Then the Fock exchange operator is evaluated as shown in Equ.~\ref{eqn:mixed_vx}. Due to the introduction of the occupation matrix $\sigma_{i,k}$, a triple loop is needed in calculating the Fock exchange operator among wavefunctions $\phi_i$, $\phi_j$ and $\phi_k$. This requires a total number of $N^3$ FFTs, leading to a computational complexity of $O(N^3N_glogN_g)$, where $N_g$ is the number of grid points and $N$ is the number of electrons. This computational complexity is higher than that in zero-temperature rt-TDDFT or ground-state calculations, which requires only $N^2$ FFTs since the occupation matrix $\sigma$ is diagonal, and only two-electron interaction between $\phi_i$ and $\phi_j$ are evaluated (as shown in Eq.~\ref{eqn:pure_vx}). 
 
In our baseline implementation,  we take the following steps to optimize the calculation of the Fock exchange operator.
\textbf{(a) Band-by-band implementation.} 
First, we implement the Fock exchange operator in a band-by-band manner, and CUFFT and FFTW are utilized on GPU and ARM platforms. The gaps between the FFT invoke are filled via CUDA customized kernels or OpenMP accelerated computations. Note that no CPU-GPU synchronization is during the calculation. 
\textbf{(b) Multi-batch implementation.} For GPU platform, we further utilize a multi-batch strategy to enhance its bandwidth utilization. Each A100 GPU has a bandwidth of 1.5 TB/s\, and the band-by-band implementation cannot fully exploit the hardware limit. To improve the performance, instead of sending the data $\Phi^*\Phi$ one by one,  we perform multi-batch operations in customed CUDA kernels, cuFFT, FFT, and MPI\_Bcast to fully saturate the memory and network bandwidth. Especially, the multi-batch implementation can also reduce the latency of CPU-GPU kernel launch. The batch size is set to 16. We find that the multi-batch implementation can greatly improve the performance of the Fock exchange operator compared to the band-by-band implementation.

\subsection{Other calculations}
As Amdahl`s law indicates, all calculations have to be optimized to achieve a desirable speedup. Thus in our PWDFT implementation, we have moved almost all calculations to the GPU and ARM cores besides the computationally intensive Fock exchange operator. 

\textbf{1. Charge density evaluation.}
The  charge density $\rho$ is calculated via $\sum\limits_{i,j=1}^{N}\phi_{i}\sigma_{ij}\phi^{\ast}_{j}$ in the PT-IM method. The introduction of occupation matrix $\sigma_{i,j}$ has increased the computational complexity of charge density calculation from $O(N^2logN_g)$ (ground state) to $O(N^3logN_g)$ due to the interaction for each $i,j$ pair of the wavefunctions. Note that the wavefunctions $\Phi$ have to be communicated across all MPI tasks via MPI\_Bcast due to the parallel distribution as detailed in Sec.~\ref{sec:paralellization_scheme}, and all calculations are evaluated either via efficient libraries such as CUFFT/FFTW or hand written CUDA kernels/OpenMP. 

\textbf{2. Anderson mixing and orthogonalization.} The Anderson mixing in PT-IM solves the least square problem for each wavefunction and $\sigma$. Note that the least square problem can be very small ($20\times20$ in our implementation), thus the main computation is the evaluation of the overlap matrix that can be efficiently calculated via grid-point parallelization. Our implementation, requires 20 copies of the wavefunctions, which can cost lots of HBM if stored in the GPU. Thus in our GPU implementation, all wavefunctions are stored on the CPU to save GPU memory footprint and then copied to GPU for matrix-matrix multiplication to obtain the overlap matrix. The orthogonalization step is also accelerated via calling efficient libraries and hand-optimized kernels. 

In summary, we develop a solid baseline version of PT-IM within the PWDFT package by incorpoarting the optimizations described above.
\section{Further optimizations}\label{sec:further_optimization}
A solid baseline version of PT-IM is implemented within the PWDFT package on the GPU and ARM platform, as detailed in Sec.~\ref{sec:implementation}. 
However, despite our efforts in optimizing almost all calculations with multi-threaded parallelism and GPU, our baseline code still encounters several challenges: Firstly, it suffers from surging computation and communication complexity introduced by the occupation matrix $\sigma$, as delineated in \eqref{eqn:mix_P(r,r')}\eqref{eqn:mixed_vx}. Paricularly, the number of FFTs in the evaluation of the Fock exchange operator grows from $N^2$ (ground state) to $N^3$ (PT-IM). Since the Fock exchange operator is required in each $H\Phi$ calculation, we will have to calculate 25 $V_x\Phi$ in each time step on average. Moreover, the communications cost can be optimized via computation-communication overlap and asynchronous communication. In this section, we will introduce how to address the challenges above through step-by-step optimizations.

\subsection{Algorithm innovation}
\subsubsection{Reduce the complexity of  Fock exchange operator and density calculation by occupation matrix diagonalization} \label{sec:digonalization}
\begin{figure}[ht]
\centering
  \includegraphics[width=0.48\textwidth]{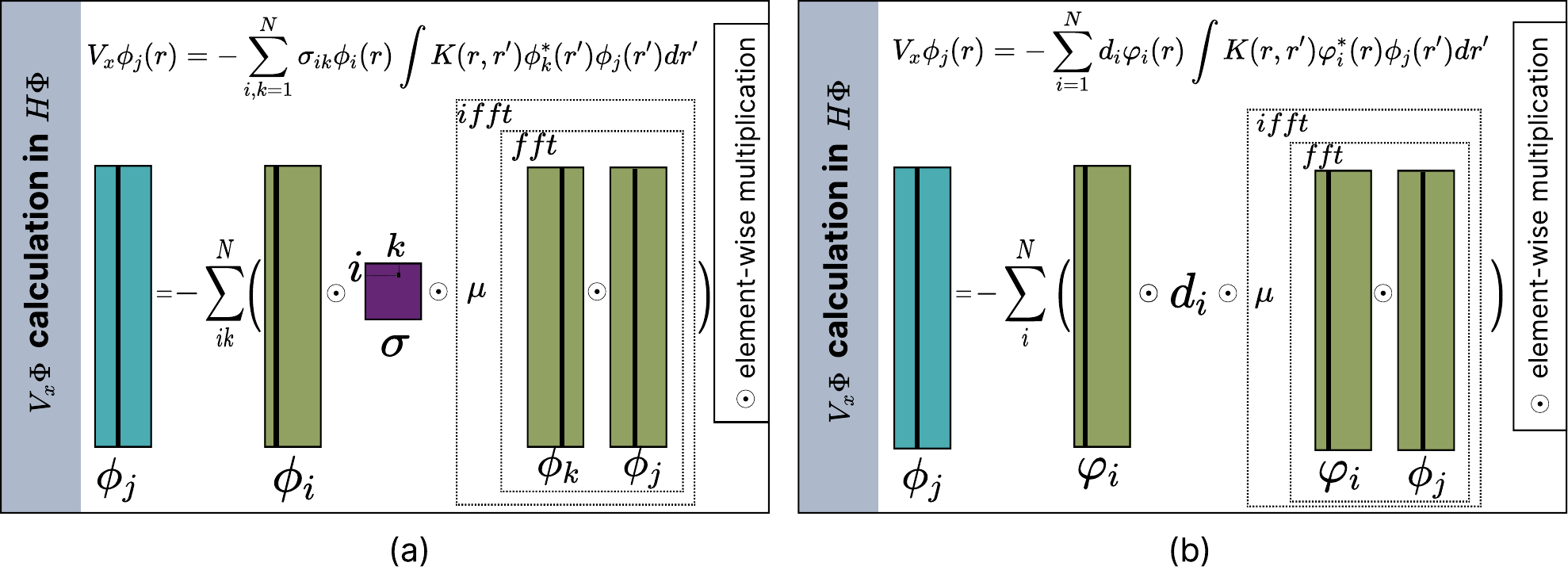}
  \caption{Evaluation of the Fock exchange operator. (a) Baseline. (b) Accelerated by diagnonalization.}
  \label{fig:diagonalization}
\end{figure}
In PT-IM, the occupation number matrix $\sigma$ introduces extra computation into the evaluation of charge density and Fock exchange operator. For example, in PT-IM the Fock exchange operator evaluation requires $10^9$ FFTs for a physical system with $10^3$ orbitals. The extensive computational cost hinders the time-to-solution of PT-IM to more than 30 minutes per time step for a physical system of 384 atoms. One key observation is that $\sigma$ is a Hermitian matrix, whose eigenvectors are orthogonal to each other. Hence, we can diagonalize it:
\begin{equation}
\sigma_{t} =QDQ^{\ast}.\label{eqn:diag_sigma}
\end{equation}
Here D is a diagonal matrix with diagnonal elements $d_{1}, d_{2}, ..., d_{N}$. 
We can set $\boldsymbol{\varphi} = \Phi Q$, so density matrix can be written as: 
\begin{equation}
P(r,r') =\sum\limits_{i}^{N}\varphi_{i}(r) d_{i}\varphi^{\ast}_{i}(r').\label{eqn:p_diag}
\end{equation}
Meanwhile, the result of $V_{x}$ applied to an orbital $\phi_{j}$ then is given by:
\begin{equation}
(V_{x}[P])\phi_{j}(r)= -\sum\limits_{i}^{N} d_{i}\varphi_{i}(r)\int K(r,r')\varphi^{\ast}_{i}(r')\phi_{j}(r')dr',\label{eqn:vxdiag}
\end{equation}
As illustrated in Fig.~\ref{fig:diagonalization}(b), the only additional overhead introduced is the single diagonalization of $\sigma$ after each update and the basis set transformation of the wavefunctions during the calculation of density and exchange operators. The number of FFTs in $V_x\Phi$ calculation is greatly reduced from $O(N^3)$ to $O(N^2)$, decreasing from a triple loop to a double loop, and communication volume from $O(N_gN^2)$ to $O(N_gN)$. Fig.~\ref{fig:diagonalization} shows the comparison between the naive and optimized versions of the Fock exchange operator. Similarly, the number of FFTs in the calculation of charge density can also be reduced from $O(N^2)$ to $O(N)$.
\subsubsection{Reduce the frequency of Fock exchange operator by adaptively compressed exchange (ACE)}\label{sec:ACE}
\begin{figure}[ht]
\centering
  \includegraphics[width=0.48\textwidth]{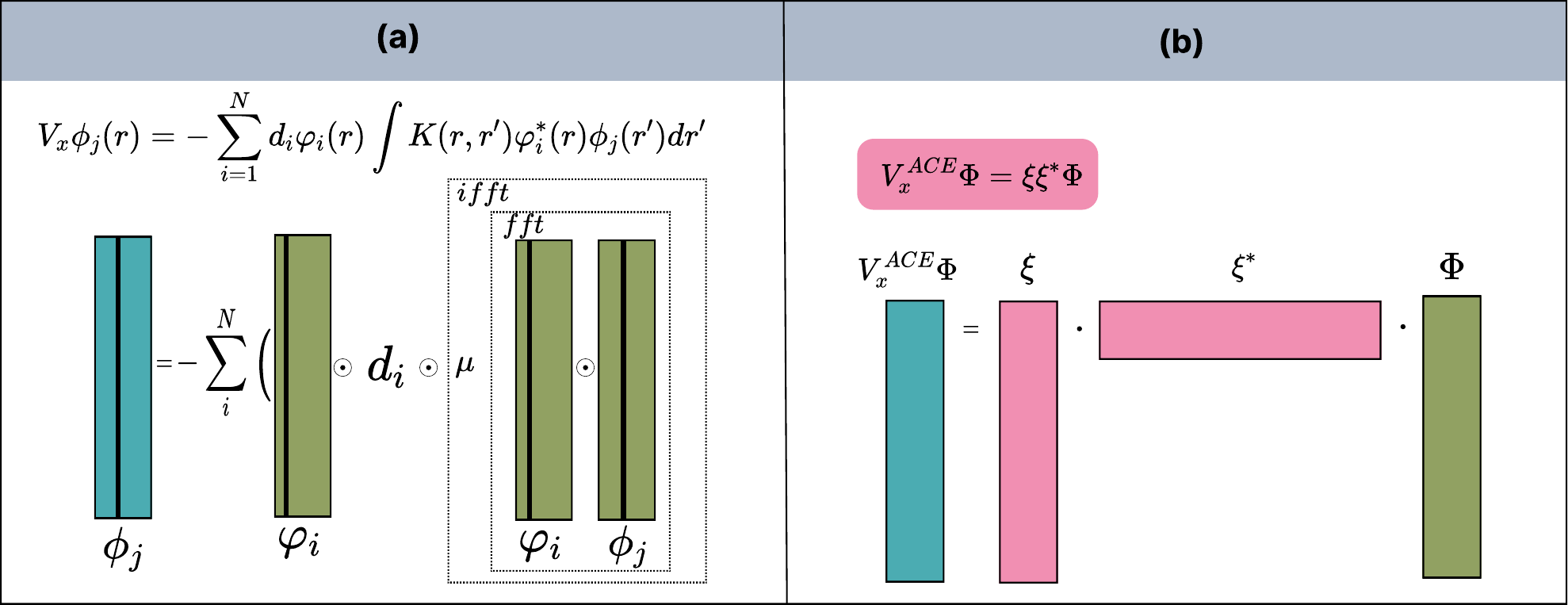}
  \caption{Evaluation of $V_x\Phi$. (a) Direct two-electron integral. (b) ACE operator.
  }
  \label{fig:ACE_optimized}
\end{figure}
The Fock exchange operator remains the most computationally intensive part after introducing the occupation matrix diagonalization, i.e., it still takes $90\%$ of the total time for a silicon system of 384 atoms. One way to further optimize it is to reduce the frequency of the Fock exchange operator by adopting the ACE formulation, which is introduced by Lin~\cite{lin2016adaptively}. The construction of the low-rank ACE operator is as follows: 
\begin{equation}
\begin{split}
W_{i}(r) = (V_{x}\phi_{i})(r) = (V^{ACE}_{X}\phi_{i})(r),\\
V^{ACE}_{X}(r,r') = -\sum\limits_{i=1}^{N_{e}}\xi_{k}(r)\xi_{k}(r').
\label{eqn:vx_ace}
\end{split}
\end{equation}
More theoretical details on $W$ and $\xi$ are described in Ref.~\cite{lin2016adaptively}. Fig.~\ref{fig:diagonalization} shows the computational procedure of both two-electron integral and ACE operator.

To integrate the ACE method into PT-IM, two ACE operators are required due to the implicit midpoint rule: $V^{ACE}_{x_n}$ and $V^{ACE}_{x_{n+\frac{1}{2}}}$. 
Those ACE operators are incorporated into PT-IM via a double self-consistent field (SCF) loop, where both $V^{ACE}_{x_n}$ and $V^{ACE}_{x_{n+\frac{1}{2}}}$ are constructed in the outer SCF. 
During the evaluation of $H\Phi$ of the inner SCF, the ACE operators $V_X^{ACE}$ can replace the previous Fock exchange operator, transforming the previously two-electron integral into more efficient matrix-matrix multiplications of size $N_g\times N$. Fig.~\ref{fig:ACE_optimized}(b) shows a detailed workflow of the PT-IM-ACE. 
\begin{figure}[ht]
\centering
  \includegraphics[width=0.48\textwidth]{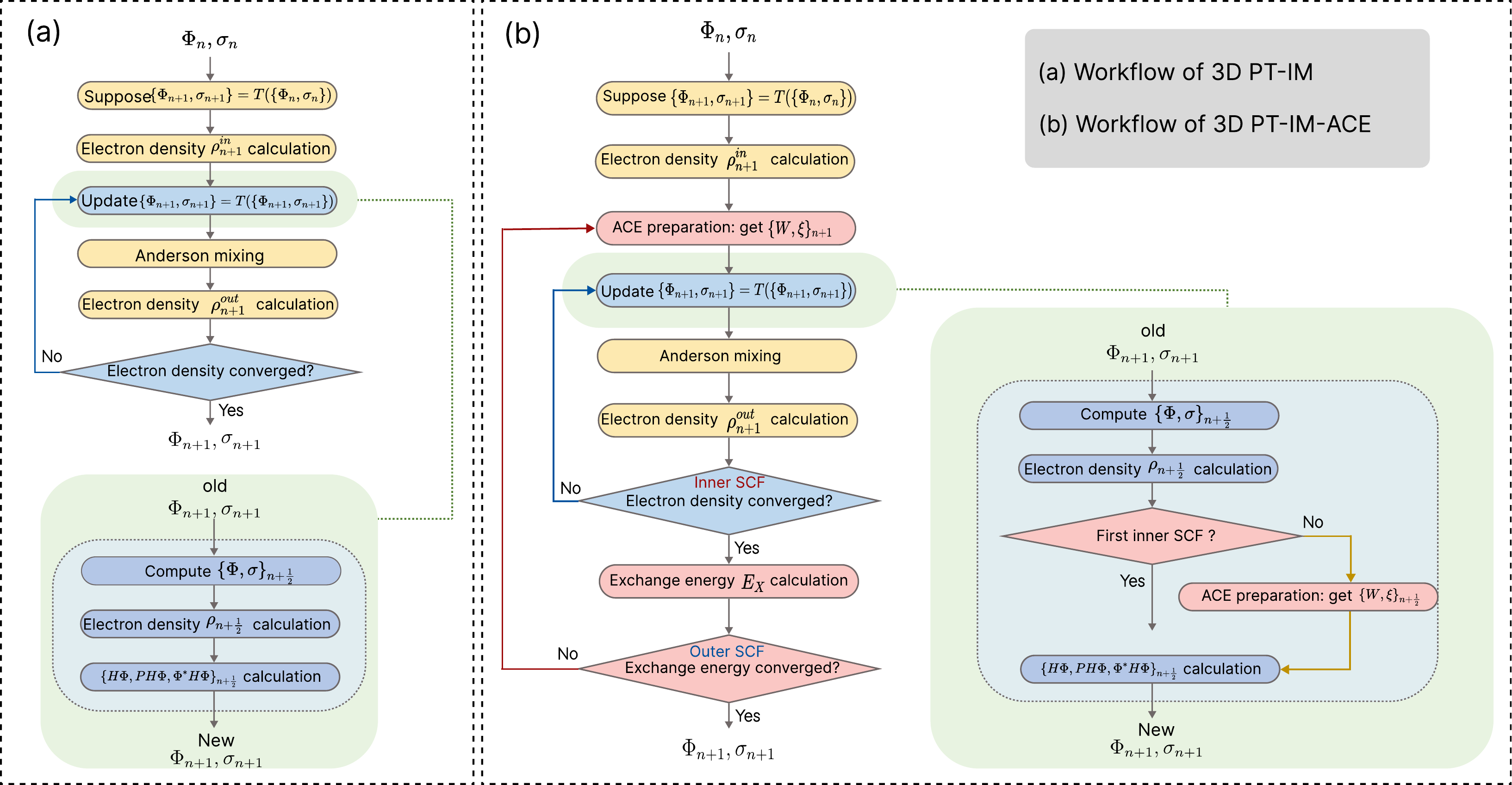}
  \caption{One time step propagation of the rt-TDDFT using (a) PT-IM (b) PT-IM-ACE with double loop to reduce the frequency of Fork exchange operator.}
  \label{fig:flowchart}
\end{figure}
Note that ACE operator can greatly reduce the frequency of the Fock exchange operator application.
For example, to fully converge for a silicon system of 384 atoms, an average of $25$ SCF steps are required, meaning that 25 Fock exchange operators are evaluated in the previous implementation (Fig.~\ref{fig:flowchart}(a)). With the introduction of ACE operator, it takes about $5$ outer SCF iterations, with each outer SCF averaging $13$ inner SCF iterations. This optimization reduces the number of  Fock exchange operator calculation by 20, or 80\%, in a single time-step propagation. 
\subsection{System innovation}\label{sec:system-innovaion}
Since all computational intensive parts have been migrated to GPU/ARM in Sec.~\ref{sec:implementation}, we focus on the communication time and memory footprint in this subsection. The most time-consuming MPI operation is the wavefunction MPI\_Bcast in the evaluation of the Fock exchange operator. Fig.~\ref{fig:communication}(a) illustrates a naive implementation of MPI\_Bcast with 4 MPI tasks. In this setup, 4 steps of MPI\_Bcast are performed to evaluate the Poisson-like equation for all wavefunction pairs (i,j). To reduce the communication time associated with the wavefunctions, we perform several steps of optimization.

\subsubsection{Ring-based point-to-point pattern} \label{sec:orbital-rotation}

\begin{figure}[ht]
\centering
  \includegraphics[width=0.48\textwidth]{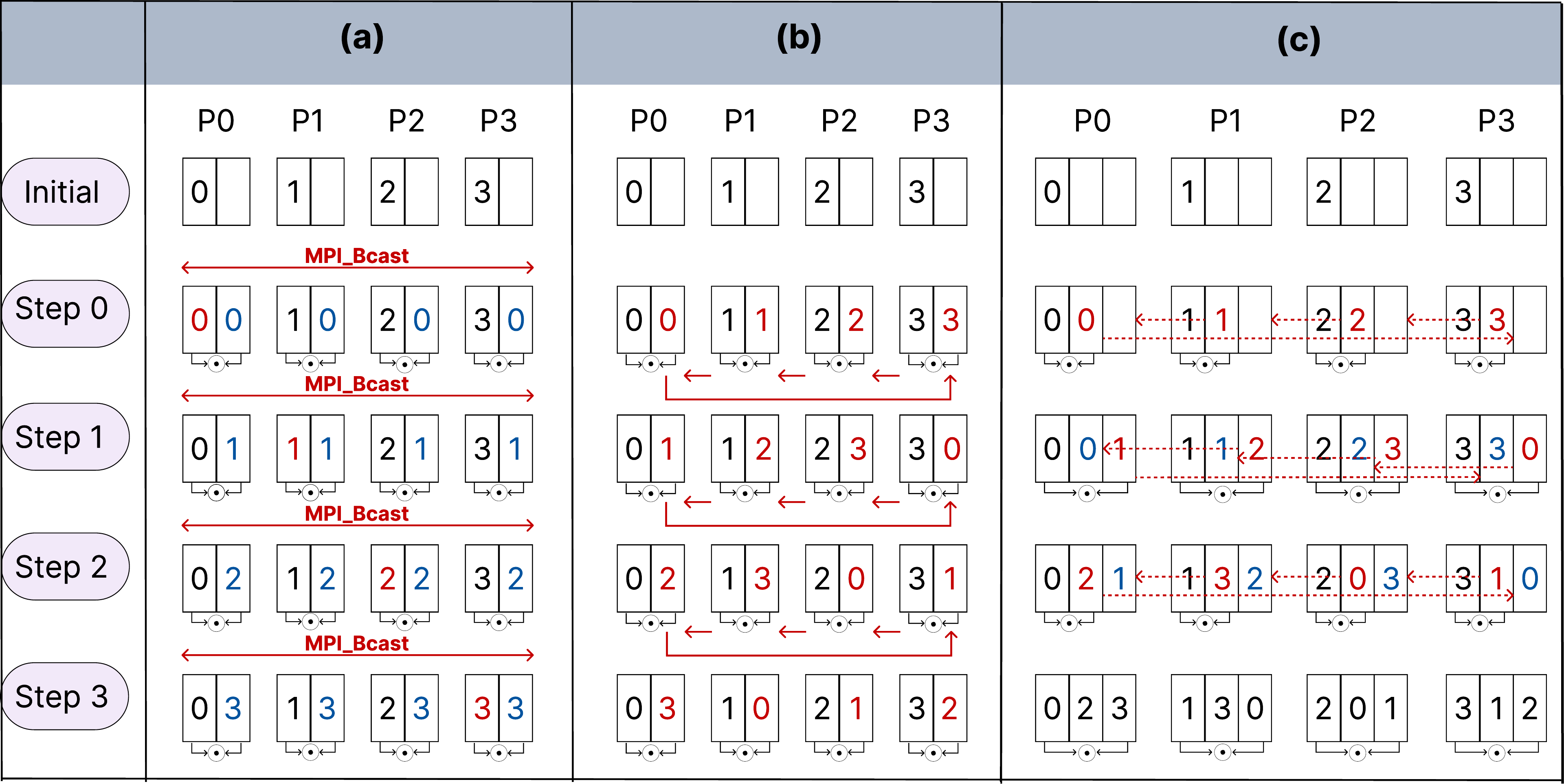}
  \caption{Communication pattern of wavefunctions across 4 processes. (a). Bcast-based method. (b). Ring-based point-to-point pattern. (c). Asynchronous ring-based method. The red two-way arrow solid line indicates MPI\_Bcast communication, and the red one-way arrow solid line is point-to-point communication. The dashed red one-way arrow stands for asynchronous point-to-point communication. $\odot$ denotes element-wise multiplication between two wavefunctions.}
  \label{fig:communication}
\end{figure}

We propose a ring-based point-to-point (p2p) communication pattern, as shown in Fig.~\ref{fig:communication}~(b). In this approach, wavefunctions are rotated among processes through point-to-point MPI communications. Within each step, MPI tasks send and receive wavefunctions from its adjacent processes. 

The ring-based method offers distinct advantages over the conventional broadcast approach in both communication pattern and latency. Unlike broadcasting, which requires global communication and can impact the entire network, the ring-based approach limits communication to neighboring processes. This localized communication significantly reduces network load and minimizes congestion. Additionally, in terms of latency, the ring-based method ensures that each communication step occurs within a single hop, which is highly advantageous in most network topologies. As a result, this method greatly improves scalability by reducing communication burdens and times.

\subsubsection{Asynchronous ring-based method}\label{sec:async-orbital-rotation}

Furthermore, the performance of the ring-based method can be substantially enhanced by leveraging asynchronous execution to overlap communication with computation. The process is shown in Fig. \ref{fig:communication}(c).
In each step, a process asynchronously transmits its local wavefunctions (or those received in the previous  step)  to the next neighboring process while simultaneously beginning to asynchronously receive wavefunctions from the previous neighbor. After completing these initial computations, the process waits for the communication phase to finish before proceeding to the next step. This iterative process consists of \textit{mpisize} steps.

Overlapping communication with computation can further reduce the total runtime, significantly improving program performance. While this technique can also be applied to Bcast-based method, its effectiveness depends on whether computation or communication takes longer. As noted in Sec. \ref{sec:comm_analysis}, our tests show that communication time exceeds computation time, meaning that communication ultimately determines the total runtime. The broadcast method generally  increases communication time, thereby reducing  the benefits of overlapping computation with communication.

\subsubsection{Reduce memory footprint using shared memory mechanism}\label{sec:SHM}
\begin{figure}[ht]
  \includegraphics[width=0.48\textwidth]{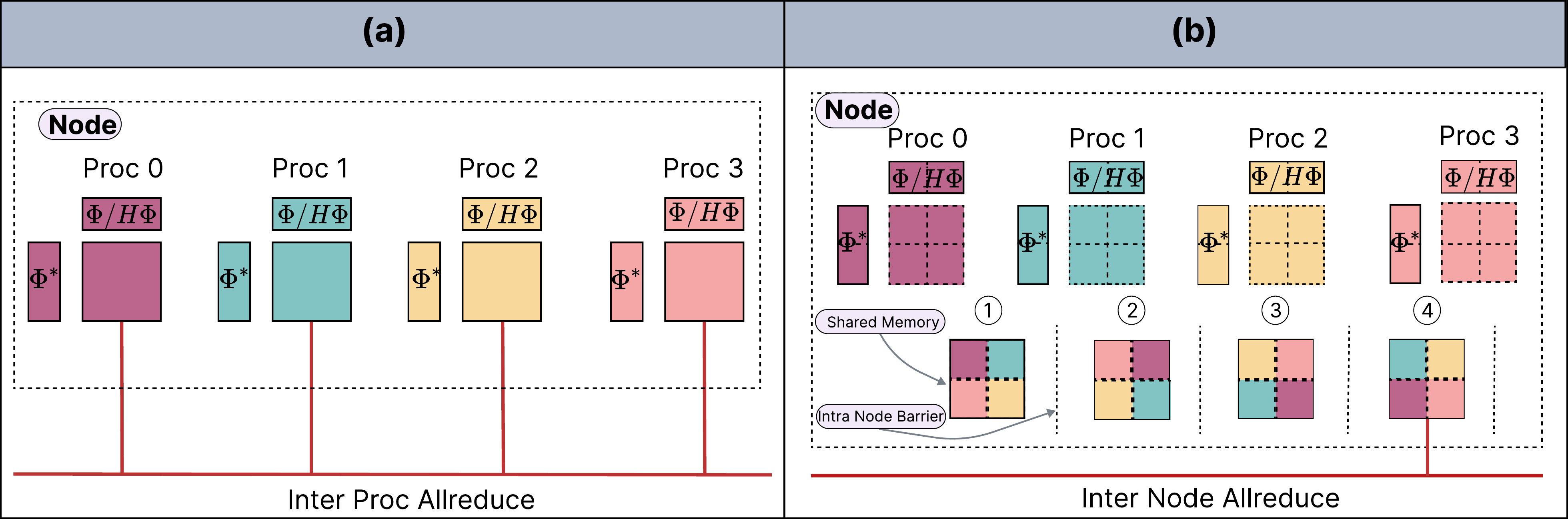}
  \caption{Calculation for $\Phi^{\ast}\Phi$ or $\Phi^{\ast}H\Phi$ across 4 processes. (a) The original. (b) optimized by shared memory mechanism. The matrix with dashed border has not been allocated memory.}
  \label{fig:shared-memory}
\end{figure}
The idea is to use inter-process shared memory to reduce memory usage. In the original implementation, certain matrices, such as $\sigma$, and  intermediate results like $\Phi^{*}\Phi$ and $\Phi^{*}H\Phi$, are not memory scalable. 
Consider a system with 768 silicon atoms and 1920 electronic orbitals, characterized by a substantial grid size of $N_g = 324000$. When using more than 168 processes, the memory advantage of scalable matrices, such as wavefunctions, diminishes. At this stage, the memory consumed by non-scalable square matrices becomes significant and cannot be ignored.

Our key idea is to use shared memory between processes to store these matrices, as they are identical across all processes. Shared memory is allocated using the MPI SHM Extension\cite{brinskiy2017introduction}, allowing processes on the same computing node to share the same matrix. If $p$ processes are launched on a node, the memory usage for these matrices is reduced to $1/p$ of the original amount. This optimization primarily enables the simulation of larger systems.

As for performance improvement, although inter-process allreduce was replaced with inter-node allreduce, reducing the number of processes involved in communication to one-quarter of the original, there was no significant performance gain. This lack of improvement is due to the introduction of remote memory access in NUMA systems. When the runtime, the $\Phi^{}\Phi$ and $\Phi^{}H\Phi$ matrices we stored in the shared memory were allocated to a single NUMA node in the physical memory, which would lead to the remote memory access problem when the computing cores in the other three NUMA nodes accessed this matrix, resulting in a loss of performance. In fact, with our method, by sacrificing little computing performance, we reduce the number of communication process, communication volume and memory footprint of this part to a quarter of the original. This enables us to scale up to a larger size.
\section{machine configuration}~\label{sec:machine}
All our tests are performed on both ARM and GPU platforms.
The first machine is Fugaku, an ARM many-core supercomputer currently ranked fourth in the Top500 list \cite{top500}, with a theoretical peak performance of 537.21 PFLOPS. Fugaku is comprised of 158,976 computing nodes interconnected through a 6D-torus network. Each node is equipped with one A64FX ARM CPU, which has four core memory groups (CMGs). Each CMG has 13 cores (1 for OS and 12 for compute) and 8GB of HBM2 memory (32GB HBM2 per node). Additionally, each computing core supports 512-bit SVE vector instructions, allowing an A64FX to reach a theoretical peak performance of 3.38 TFLOPS at 2.2GHz, with a theoretical memory bandwidth of 1024GB/s.

The second platform is a GPU cluster featuring NVIDIA A100 GPUs. Each computing node is equipped with one ARM-based Kunpeng-920 CPU, 256GB of DDR4 memory, and 4 NVIDIA A100 GPUs.  Each Kunpeng-920 CPU has 128 cores distributed across four NUMA domains, each supporting 128-bit NEON vector instructions. Each A100 GPU accelerator offers a theoretical peak performance of 9.7 TFLOPS (19.5 TFLOPS with tensor cores) and 40GB of HBM2 memory, achieving a theoretical bandwidth of 1.5TB/s. The CPU and GPUs are interconnected via a PCIe bus with a bi-directional bandwidth of 64GB/s.  The computing nodes are interconnected through a fat-tree network.

\section{physical system}~\label{sec:physical_system}
Silicon systems ranging from 48 to 3072 atoms, corresponding to the supercell constructed from 1$\times$1$\times$3 to $6\times8\times8$ unit cells. Each simple cubic unit cell consists of 8 silicon atoms with the lattice constant being 5.43 Å. 
In our accuracy tests, the number of extra states is set to $N_{atom}$, and it is set to $\frac{1}{2} N_{atom}$ in all other tests.

In our tests, the external potential is a laser pulse shown in Fig. \ref{fig:accuracy}(a), and its wavelength is 380 nm. The total simulation time is 30 fs, with a time step of 50 as for both PT-IM and PT-IM-ACE methods. The stopping criteria is set to $1.0 \times 10^{-6}$ for  electron density and  exchange energy errors. We use the SG15 Optimized Norm-Conserving Vanderbilt (ONCV) pseudopotentials\cite{schlipf2015optimization,hamann2013optimized} and HSE06 functionals\cite{heyd2003hybrid} in all tests. The kinetic energy cutoff is set to 10 Hartree and the temperature is set to 8000K. The average number of outer and inner SCFs is 5 and 13, respectively. The maximum Anderson mixing dimension is set to 20.

For the system with 1536 atoms, the number of grid points for a wavefunction is $N_g = 60 \times 90\times 120 = 648, 000$. This corresponds to a charge density grid $120 \times 180 \times 240$. The Fock exchange operator is evaluated on the wavefunction grid. The number of orbitals is $N = 1536 \times 2 + \frac{1}{2} \times 1536 = 3840$.

\section{Physical results }\label{sec:physical_results}

In this section, we test the accuracy of the optimized code and descibe the motion of electrons during the rt-TDDFT simulation.
\subsection{Accuracy}
\begin{figure}[ht]
  \includegraphics[width=0.48\textwidth]{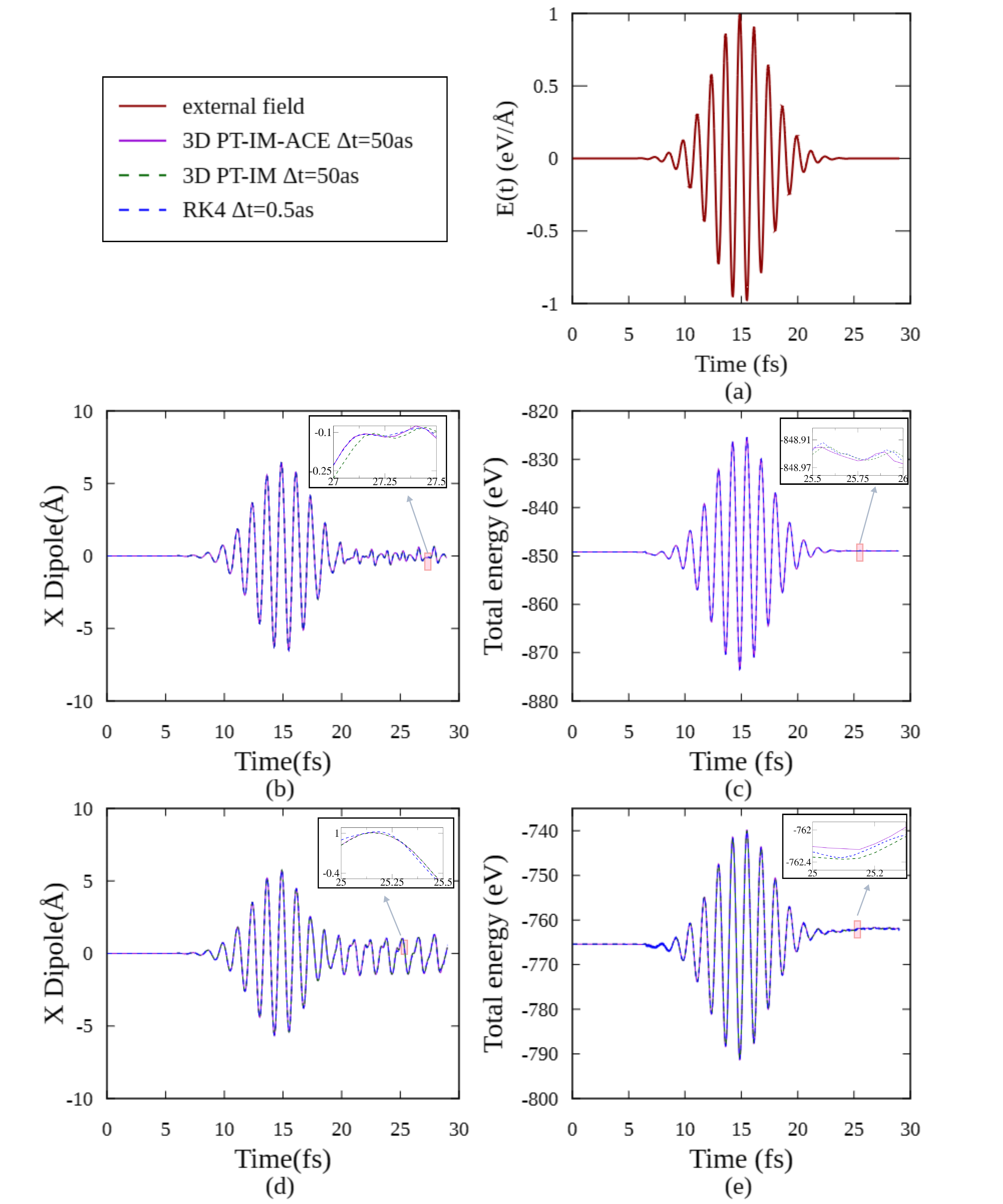}
  \caption{Electron dynamics of an 8 atom silicon system under a laser pulse with 380 nm. (a) Electric field along the x direction. (b) Dipole moment along the x direction in pure states. (c) Total energy in pure states. (d) Dipole moment along the x direction in mixed states(Total states = 24). (e) Total energy in mixed states(Total states = 24).}
  \label{fig:accuracy}
\end{figure}

The accuracy is evaluated using the dipole moment along the x-direction and the total energy, as shown in Fig.\ref{fig:accuracy}. For the 380 nm laser case, Fig.\ref{fig:accuracy} demonstrates that the results of PT-IM-ACE with a 50 as time step fully match those obtained using the RK4 method with a time step 100 times smaller. Furthermore, the enlarged section in Fig.~\ref{fig:accuracy} confirms that PT-IM-ACE provides a very good approximation to the electron dynamics compared to RK4 during the final 100 time steps (25-30 fs), regardless of whether the system is in a pure or mixed state. It is important to note that electrons already exhibit fractional occupation at the beginning of the finite temperature (8000K) rt-TDDFT simulation.
\subsection{Electrons motions}
\begin{figure}[ht]
  \includegraphics[width=0.48\textwidth]{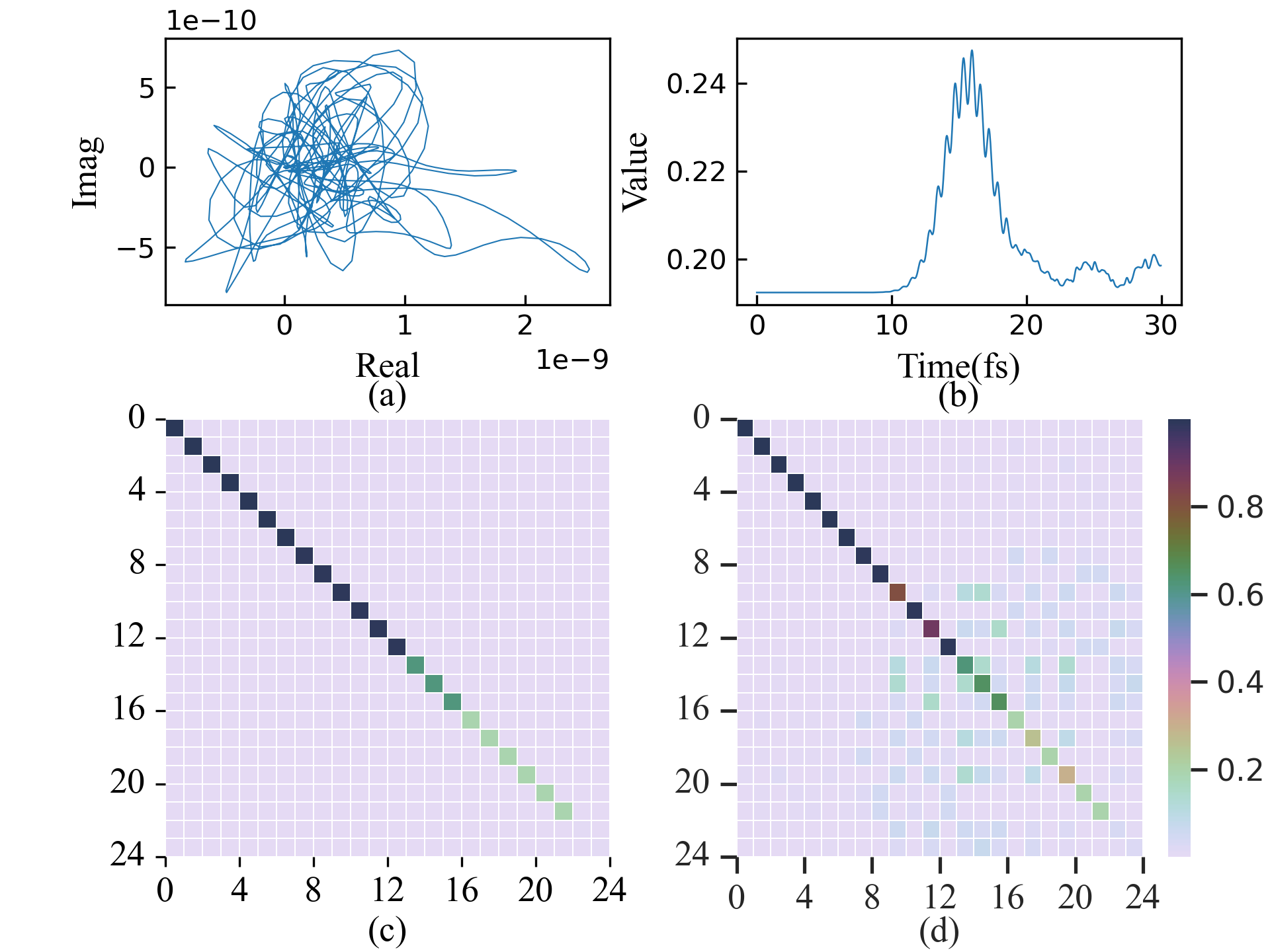}
  \caption{States evolution of an 8-atom silicon system under laser pulse irradiation over 30 fs with 4 processes. (a) Relationship between the real and imaginary parts of the off-diagonal element $\sigma_{t}(0,2)$ over 30 fs. (b) Variation of the diagonal element $\sigma_{t}(22,22)$ over 30 fs. (c) Initial $\sigma_{t}$. (d) Final $\sigma_{t}$. }
  \label{fig:sigmat}
\end{figure}
The motion of electrons in finite temperature rt-TDDFT is shown in Fig.\ref{fig:sigmat}, with the laser pulse shown in Fig.~\ref{fig:accuracy}(a). Initially, the occupation number matrix $\sigma_t$ is depicted in Fig.\ref{fig:sigmat}(c), with elements from 0 to 1 indicating the probability of electron occupying each orbital. During the simulation, the variation of the off-diagnal elemment $\sigma_{t} (0,2)$ over time is shown in Fig.\ref{fig:sigmat}(a), demonstrating the stochastic nature of electron motion. Meanwhile, as an example of diagonal elements, the variation of $\sigma_{t} (22,22)$ over time is shown in Fig.\ref{fig:sigmat}(b), increasing as the external field is strengthening (10-15 fs). This indicates that the stronger the external laser field, the more active the electrons are. Enhanced electron activity in laser fields implies that modifying material's electronic structures and band properties can  significantly affect their optical responses, offering new avenues for optoelectronic device innovation.
\section{performance results and analysis}\label{sec:performance_results}

In this section, we perform detailed performance tests, including step-by-step performance improvements, strong scaling, weak scaling, and communication analysis. 
On the ARM platform, all our tests are conducted with four MPI ranks per node, matching the A64FX's four NUMA architectures. Each process launches 12 threads to manage the 12 computing cores within a CMG, with  access to 8GB of memory. On the GPU platform, each compute node is equipped with four A100 GPUs, so we initiate four MPI ranks per node, with each process controlling one A100 GPU. Consequently, each process has access to 64GB of host memory and 40GB of GPU memory.

\subsection{Step-by-step performance improvement} \label{sec:step_by_step}

\begin{figure}[ht]
 \centering
  \includegraphics[width=0.48\textwidth]{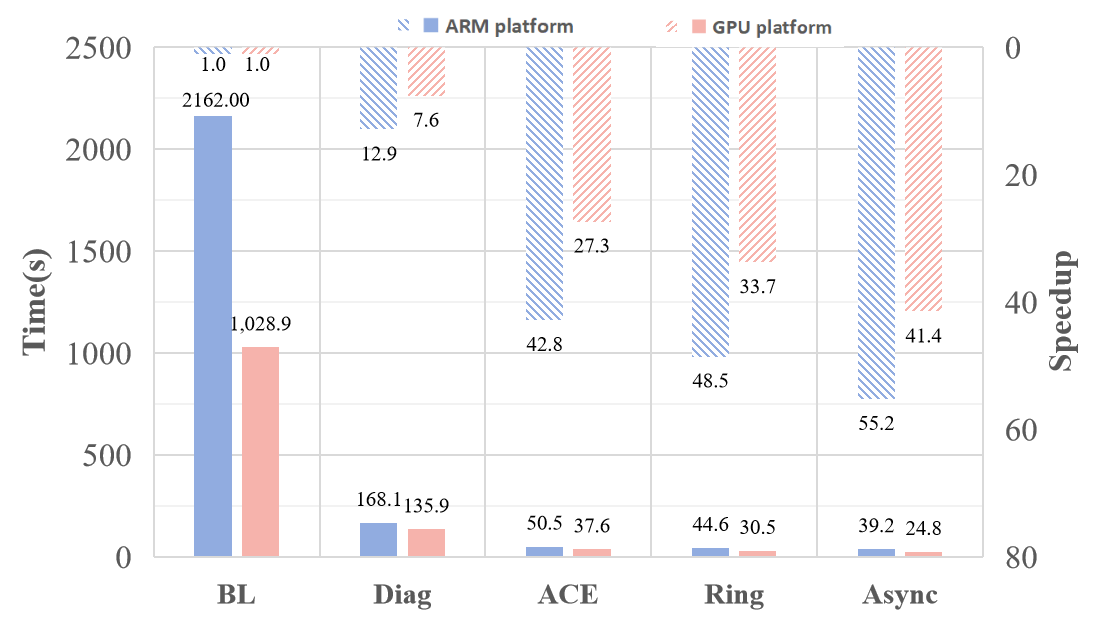}
  \caption{Step-by-step performance improvement for one time step on GPU/ARM platforms with 384 silicon atom system using 240/24 nodes on ARM/GPU platform. The baseline is the results calculated by the initial PTIM method accelerated by OpenMP and GPU.}
  \label{fig:step_by_step}
\end{figure}

On both platforms, step-by-step performance improvements are evaluated using a 384 silicon atom system on 240 ARM nodes and 24 GPU nodes. Results are shown in Fig. \ref{fig:step_by_step}. The baseline test (BL) represents the original PTIM algorithm, accelerated by OpenMP and GPU on two respective platforms.

\subsubsection{Occupation matrix diagonalization}
As shown in Fig.~\ref{fig:step_by_step}, the occupation matrix diagonalization method, labeled as "Diag" in the figures, is introduced in Sec. \ref{sec:digonalization}. For the 384 atom system, this method accelerated per step performance of the PT-IM in PWDFT by 12.86x on the ARM and 7.57x on the GPU platform. In Sec.\ref{sec:digonalization}, we describe the diagonalization algorithm, which reduces the computational complexity of $V_{x}\Phi$. 
As system size $N$ increases, the importance of $V_{x}\Phi$ calculation grows, and the benefit of complexity reduction becomes more significant. Therefore, the speedup of the diagonalization of occupation matrix is more substantial for larger systems.
\subsubsection{ACE method}
The ACE operator method,  introduced in Sec.\ref{sec:ACE} and labeled as "ACE" in Fig. \ref{fig:step_by_step}, reduces the number of computations for $V_x\Phi$ from 25 to 5. On the ARM/GPU platform, for the 384 atom system, the computation time of $H\Phi$ decrease from 148.5s/110.6s to 6s/20.3s, with the total ACE preparation time being 23s/17.4s. As shown in Fig.~\ref{fig:step_by_step}, the ACE operator  accelerates the per-step time by 3.3x/3.6x.

\subsubsection{Ring-based method}
The performance gain from the ring-based method, detailed in Sec.~\ref{sec:orbital-rotation} and denoted as a "Ring", is shown in Fig. \ref{fig:step_by_step}. Compared to the Bcast-based method in the previous step, it  accelerates the 384 atom system by 1.13x/1.23x on the ARM/GPU platform.

\subsubsection{Asynchronous ring-based method}
The asynchronous ring-based method is detailed in Sec. \ref{sec:async-orbital-rotation} and is denoted as "Async" in Fig. \ref{fig:step_by_step}. Compared to the ring-based method in the previous step, it gains a speedup of 1.14x/1.23x on the ARM/GPU platform for the 384 atom system. 

Further details on the MPI communication time across different methods will be analyzed in more detail in Sec. \ref{sec:comm_analysis}. We remark that the more nodes used, the greater the improvement of communication optimization.
\subsection{Strong scaling} \label{sec:strong_scaling}

\begin{figure}[ht]
  \centering
\includegraphics[width=0.48\textwidth]{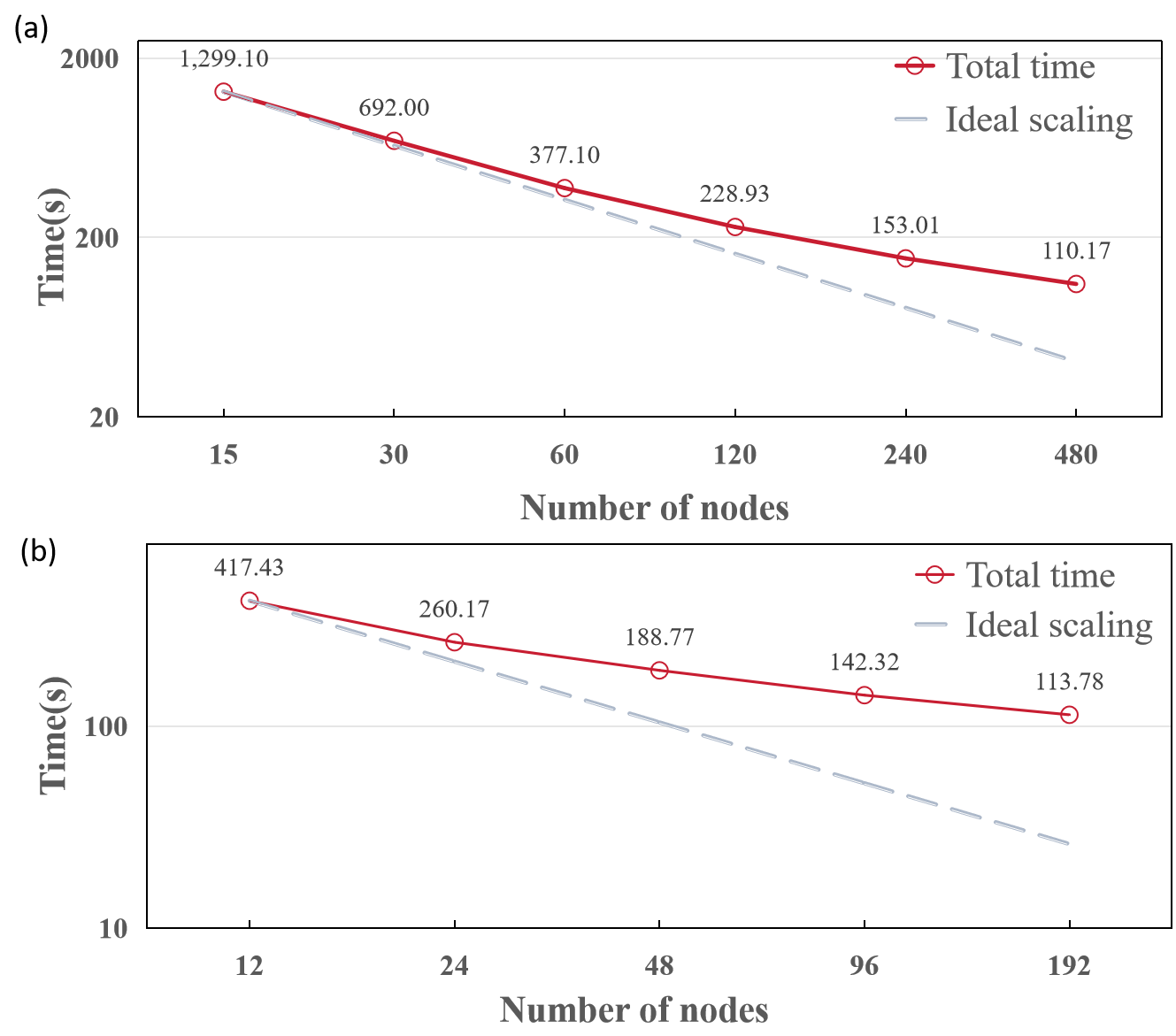}
  \caption{Strong scaling: wall clock time per 50 as for silicon systems on two HPC systems. The “ideal scaling” here scales as $O(N)$. (a). Strong scaling with 768 silicon atom system on ARM platform. (b). Strong scaling with 1536 silicon atom system on GPU platform.}
  \label{fig:strong_scaling}
\end{figure}
\begin{figure}[ht]
  \centering
  \includegraphics[width=0.48\textwidth]{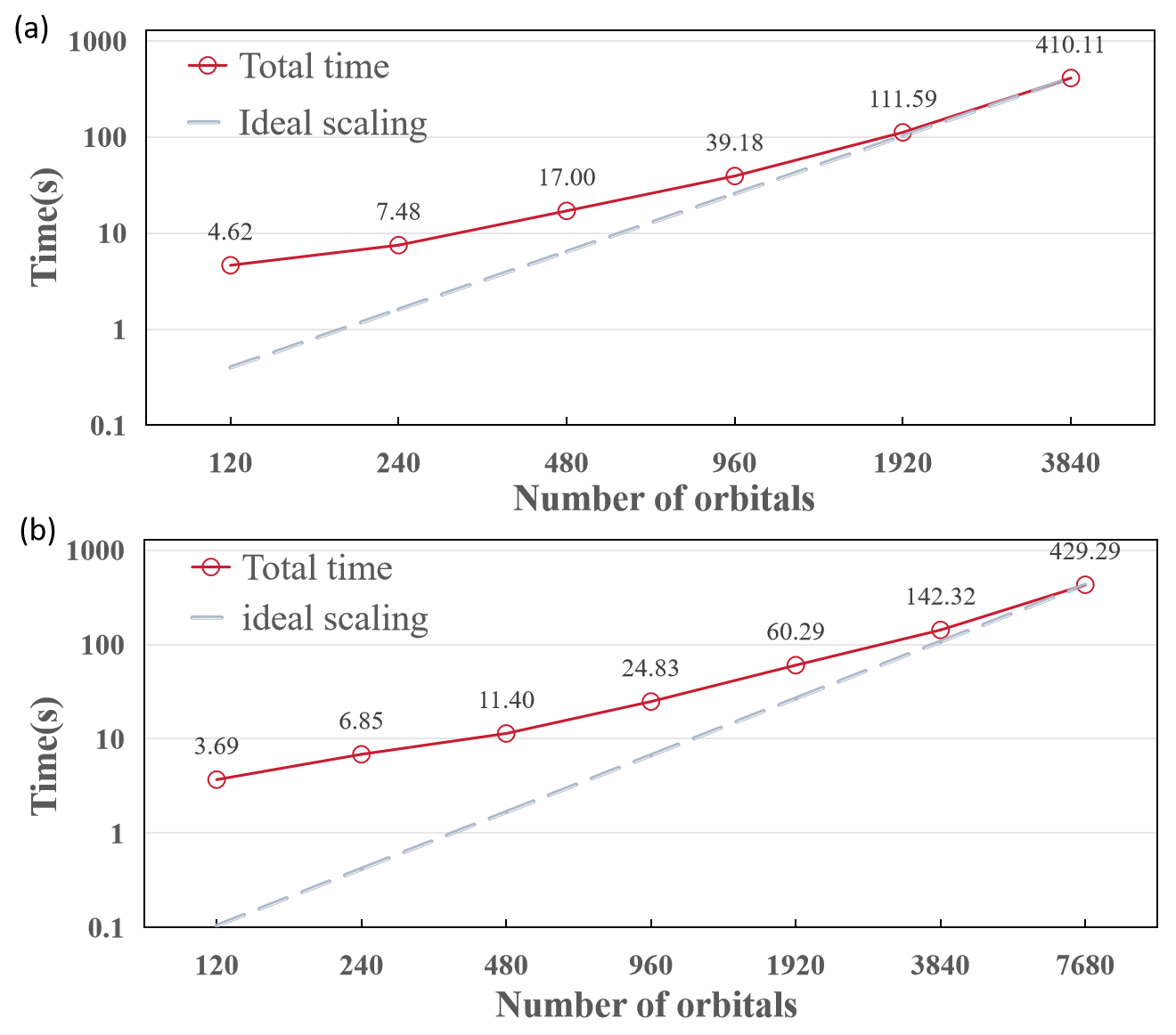}
  \caption{Weak scaling: wall clock time per 50 as for silicon systems on two HPC systems. The “ideal scaling” here scales as $O(N^2)$ on both platforms. (a).Weak scaling on the ARM platform, from 48 atoms to 1536 atoms. The number of nodes used is always set to 1/4 of the number of total orbitals in the calculation. (b).Weak scaling on the GPU platform, from 48 atoms to 3072 atoms. The number of nodes is used are always set to 1/40 of the number of total orbitals in the calculation. }
  \label{fig:weak_scaling}
\end{figure}
Fig.~\ref{fig:strong_scaling} (a)(b) shows the strong scaling of one time step using the optimized PT-IM method for a 768-atom silicon system on the ARM platform and a 1536-atom system on the GPU platform, respectively. 
On the ARM platform,  the parallel efficiency is 36.8\% when increasing the number of nodes by 32 times. On the GPU platform, the same increase yields a parallel efficiency of 22.9\%. 
The dropping of the parallel efficiency are highly related to the data communication and computational efficiency. First, as the number of nodes increases, communication time grows due to MPI\_Sendrecv in ring-based method and MPI\_Allreduce operations required by the Rayleigh-Ritz procedure. For example, on the ARM platform, when increasing from 15 nodes to 480 nodes, the MPI\_Sendrecv grows by a factor of 1.5, from 4.7 seconds to 7.1 seconds and the MPI\_Allreduce grows by a factor of 1.4, from 2.6 seconds to 3.7 seconds. On the GPU platform, when increasing from 12 nodes to 192 nodes, the MPI\_Sendrecv grows by a factor of 1.6, from 6.3 seconds to 10.1 seconds and the MPI\_Allreduce grows by a factor of 1.5, from 2.9 seconds to 4.3 seconds.
Second, the computational efficiency decreases as per-node workload scales. For example, on the ARM platform, when computing resources expand by 32 times, the computing efficiency drops to 40\% of the original. On the GPU platform, when computing resources expand by 16 times, the computing efficiency reduces to 26\% of the original.

Compared to the GPU platform, the optimized PWDFT demonstrates higher parallel efficiency on the ARM platform. This is primarily due to two factors. First, the ratio of theoretical peak performance to peak bandwidth is lower on the ARM platform (3.4 Flop/Byte) compared to the GPU platform (6.5 Flop/Byte), allowing for better performance since PWDFT is bandwidth-bound. Second, the ARM platform features a 6D torus network architecture, which provides superior network performance.
\subsection{Weak scaling} \label{sec:weak_scaling}
\begin{table*}[ht]
\caption{MPI communication time with the optimized methods for a big system of 1536 silicon atoms on both platforms}
\centering
\scalebox{0.98}{
\begin{tabular}{|cccccccccc|}
\hline
                                                       &                               & Alltoallv (s) & Sendrecv (s) & Wait (s) & Allgatherv (s) & Allreduce (s) & Bcast (s) & \thead{ Total communication \\ time(s)} & \thead{Communication \\ ratio(\%)} \\ \hline
\multicolumn{1}{|c|}{\multirow{4}{*}{\textbf{\thead{ARM \\ platform}}}} & \multicolumn{1}{c|}{ACE}       & 9.04                &       -             &        -        & 0.17                 & 14.19               & 67.22           & 90.62    & 18.92     \\
\multicolumn{1}{|c|}{}                                  & \multicolumn{1}{c|}{Ring} & 9.03                & 30.1               &       -         & 0.17                 & 14.21               & 0.03            & 53.54     & 12.73     \\
\multicolumn{1}{|c|}{}                                  & \multicolumn{1}{c|}{Async}    & 9.18                &         -           & 20.13          & 0.17                 & 14.18               & 0.03            & 43.69     & 10.65     \\ \cline{1-2}
\multicolumn{1}{|c|}{\multirow{4}{*}{\textbf{\thead{GPU \\ platform}}}} & \multicolumn{1}{c|}{ACE}       & 7.95                &     -               &       -         & 0.47                 & 4.99                & 64.85           & 78.26   & 25.72       \\
\multicolumn{1}{|c|}{}                                  & \multicolumn{1}{c|}{Ring} & 7.35                & 20.54              &      -          & 0.47                 & 4.46                & 0.89            & 33.71       & 21.13   \\
\multicolumn{1}{|c|}{}                                  & \multicolumn{1}{c|}{Async}    & 7.64                &     -               & 10.1           & 0.47                 & 4.28                & 0.82            & 23.31      & 16.38   \\ \hline
\end{tabular}
\label{tabl:communication_time}
}
\end{table*}

Fig.~\ref {fig:weak_scaling} shows the weak scaling of the optimized PT-IM code on both ARM and GPU platforms. We find that the system size is primarily constrained by the memory capacity of the hardware. For example, our optimized PWDFT can  accommodate only 1536 atoms on 960 computing nodes on Fugaku, limited by the 8GB memory capacity of each NUMA node. 
On the GPU platform, due to the number of nodes we can access, PWDFT can only scale up to 3072 atoms on 192 nodes.
Even with more nodes, the current machine configuration is unable to handle 6144 silicon atoms because of memory limitations. The simulation of 3072 atoms already consumes over 80\% of the available global memory and 75\% theoretical peak bandwidth per process.
This also indicates that PT-IM is a memory-bandwidth bounded problem. 
If our GPUs have larger global memory, it might be possible to double the simulation scale.

The computational complexity of hybrid functional rt-TDDFT simulation scales as $O(N^3)$. Notably,  on both platforms, when the number of orbitals is relatively low, doubling it results in a much smaller increase in computing time than the theoretical fourfold. However, as the system size grows, the time required to double the computational workload approaches to the theoretical four times increase. This is because, in smaller systems, less time is spent on the Fork exchange operator $V_x\Phi$. As the system scale scales up, the relative time spent on these operations increases,  eventually becoming the dominant factor in the overall simulation time.
For a smaller system with 192 atoms, simulating one time step on the GPU platform using 12 nodes takes 11.40 seconds, meaning that each femtosecond of simulation requires approximately 3.5 minutes. For a larger system with 3072 atoms, simulating one time step with 192 computing nodes takes 429.29 seconds, implying that each femtosecond of simulation takes about 2.5 hours.
\subsection{Communication analysis} \label{sec:comm_analysis}
In this section, we evaluate the MPI communication time after  system optimizations.  Table.~\ref{tabl:communication_time} shows results from 1536-atom tests on ARM and GPU platforms, using 960 and 96 nodes, respectively.
A '-' indicates that no such communication occurrs in the program.

This subsection focuses on communication, and the optimization methods prior to ACE do notdirectly reduce communication. Therefore, we start communication optimization analysis from PT-IM after ACE optimization.

On the ARM/GPU platform, the communication time for wavefunctions using the bcast-based method accounts for 74\%/83\% of the total MPI communication time. 
The ring-based method reduces the corresponding time from 67.22s/64.85s to 30.1s/20.54s.
For asynchronous ring communication, after overlapping computation and communication, the time spent on MPI\_Wait is 20.13/10.1 seconds on the ARM/GPU platforms, respectively. 

We note that on two platforms, after asynchronous communication, MPI\_Wait time is greater than zero, indicating that communication time exceeds computation time. Even with the overlap of computation and communication, communication remains the bottleneck, higlighting the importance of replacing the bcast-based method with the ring-based mechanism.

In Table. \ref{tabl:communication_time}, the proportion of communication time on the GPU platform is higher than that on the ARM platform, even though the number of processes on the GPU platform is  only one tenth of that on the ARM platform. This can be attributed to two factors. First, although the computational workload per process on the GPU is ten times that on the ARM platform, the computational power per process on the GPU platform ($9.7$ TFLOPS) is $11.5$ times greater than the ARM platform ($0.84$ TFLOPS). Second, our GPU cluster is not equipped with NVLink and does not support GPUDirect communication, which negatively impacts communication time. On the GPU platforms equipped with NVLink, such as Summit, the communication performance of our program will be further improved~\cite{elwasif2023application}.
\section{Conclusion}\label{sec:conclusion}
In this paper, we first implement a three-dimensional PT-IM algorithm in PWDFT, enabling finite-temperature rt-TDDFT simulation with hybrid functional. In terms of algorithms, we propose a diagonalization method to reduce computation and communication complexity, nearing pure state levels. Additionally, we employ the ACE method to significantly reduce the frequency of the most expensive Fock exchange operator. In terms of computer architecture, the ring-based method is utilized to optimize communication patterns and extensively reduce the communication load. Performance is further enhanced by overlapping computation and communication. The shared memory mechanism is used to alleviate the memory consumption. the correctness of our implementation is proved in the physical result. The step-by-step performance improvement test reveals our optimization achieving speeds up of 51.15 and 41.44 times speed up on ARM and GPU platforms, respectively. The strong scaling results show that when nodes scale 32 and 16 times, time-to-solution was accelerated by 11.79x/3.67X, on ARM and GPU platforms, respectively. The simulation system is respectively extended to 1536/3072 atoms(6144/12288 electrons) on two platforms. Our work paves the way for large-scale rt-TDDFT simulation for real material with finite temperatures.


\bibliographystyle{./IEEEtran.bst}
\bibliography{./Reference}
\end{document}